%% file: sample-sigconf.tex
\documentclass[sigconf]{acmart}

\usepackage{booktabs} 
\usepackage[bottom]{footmisc}
\usepackage{natbib}
\usepackage{pgfplots}
\pgfplotsset{compat=1.10}
\usepgfplotslibrary{statistics}
\usepackage{tikz}
\usetikzlibrary{positioning}
\usepackage[ruled]{algorithm2e} 

\setcopyright{none}

\acmConference[Accepted in FIRE'17 Track]{ACM FIRE conference}{Bangalore}{India} 
\acmYear{}

\begin{document}
\title{Automatic Identification and Ranking of Emergency Aids in Social Media Macro Community}

\author{Bhaskar Gautam}
\orcid{0000-0002-3258-9927}
\affiliation{%
  \institution{Data Informatics and Analytics Lab\\National Institute of Technology, Karnataka}
  \streetaddress{P.O. Srinivasnagar}
  \city{Surathkal-575025} 
  \state{India} 
  \postcode{575025}
}
\email{bhaskar.gautam2494@gmail.com}

\author{Annappa Basava}
\affiliation{%
  \institution{Data Informatics and Analytics Lab\\National Institute of Technology, Karnataka}
  \streetaddress{P.O. Srinivasnagar}
  \city{Surathkal-575025} 
  \state{India} 
  \postcode{575025}
}
\email{annappa@ieee.org}

\renewcommand{\shortauthors}{Bhaskar Gautam and Annappa Basava}

\begin{abstract}
Online social microblogging platforms including Twitter are increasingly used for aiding relief operations during disaster events. During most of the calamities that can be natural disasters or even armed attacks, non-governmental organizations look for critical information about resources to support effected people. Despite the recent advancement of natural language processing with deep neural networks, retrieval and ranking of short text becomes a challenging task because a lot of conversational and sympathy content merged with the critical information. In this paper, we address the problem of categorical information retrieval and ranking of most relevance information while considering the presence of short-text and multilingual languages that arise during such events. Our proposed model is based on the formation of embedding vector with the help of textual and statistical preprocessing, and finally, entire training 2,100,000 vectors were normalized using feed-forward neural network for need and availability tweets. Another important contribution of this paper lies in novel weighted Ranking Key algorithm based on top five general terms to rank the classified tweets in most relevance with classification. Lastly, we test our model on Nepal Earthquake dataset (contains short text and multilingual language tweets) and achieved 6.81\% of mean average precision on 5,250,000 unlabeled embedding vectors of disaster relief tweets.
\end{abstract}

%
%



\maketitle

\input{samplebody-conf}

\bibliographystyle{ACM-Reference-Format}
\bibliography{sample-bibliography} 

\end{document}

%% file: samplebody-conf.tex
\section{Introduction}
The branch of Information-driven and computational social science research is growing in importance \cite{16}, and with this trend, there is a need of common fully-automatic information retrieval model to facilitate a robust and reproducible research environment \cite{17}. Initially, Online Social Networks (OSNs) aims to allow users to communicate, connect with others and share content \cite{19} but nowadays the OSNs allows its users to broadcast personal thoughts and content. During natural or man-made disasters, these social networks provide the rich source of near-real-time data streams which helps in the formation of emergency situation summarization \cite{22}. However, increasingly usage of these social media tweets for aiding relief operations during disaster events such critical information is usually submerged with a lot of conversational content, such as sympathy for the victims of the disaster. Hence, automated Information Retrieval techniques are needed to extract availability and absence of any kind of resource to save life. \newline 

In this paper, we focus on automatic classification and ranking problem \cite{8,9} for two types of critical information, named as \textit{Need-tweets} and \textit{Availability-tweets}. Our goal is to improve classification methods and return the tweet in most relevance order with the class while considering the fact of the presence of short-text and multilingual language in OSNs. A glimpse of the challenge described in FIRE 2017 Information Retrieval from Microblogs during Disasters Track \footnote{https://sites.google.com/site/irmidisfire2017/} and also discussed in \cite{1}. Training and evaluation of the entire model is done using 70,000 tweets which were posted during the 2015 Nepal earthquake. However, including the need and availability tweets corpus also include some redundancy such as the presence of similar tweets and all such false tweets were removed by Track organizers as discussed in \cite{7} in order to preserve uniqueness of tweet labels. The purpose of this paper is to consider the highlighted problem, and the trade-offs that they entail. Our paper makes the following contributions (1) We describe the methodology and (2) release the pseudo code in order to replicate the results of entire paper. The remainder of this paper is structured as follows: in Section 2, we present the Feature resolution methods used to create a feature vector for neural network model. Section 3 presents methodology followed in the formation of entire information retrieval model. Section 4 discusses the task required to form classification model. Section 5 discusses the weighted ranking key algorithm and Section 6 depicts about the evaluation and result of the model while comparing with another team. Finally the paper concludes in Section 7 with discussing some of the conceivable use cases.

\section{Feature Resolution Methods}
In this section, we discuss the Feature Resolution Methods which helps in the formation of embedding vectors proposed to consumed by our classification model. The methods are - \textbf{Textual preprocessing} and \textbf{Statistical Preprocessing} which exploits the public information of social network users with the help of Stanford NLP Toolkit \footnote{https://nlp.stanford.edu} while preserving privacy. A detailed description of each method is elaborated in following sections.

\subsection{Textual Preprocessing}
The goal of this phase is to make all tweets in a particular semantics some of them are discussed in \cite{3, 18} considering the tweet which are written in a language (specifically English, Hindi, Nepal, and language which is the combination of Hindi + English). Tweets are represented as vectors of features, each feature being a word unigram, by performing standard text preprocessing operations whose highlights are depicted as follows:
\begin{itemize}
\item Removing non-ASCII characters.
\item Removing all the stopwords (also for Hindi and Nepali).
\item Normalizing all the text of tweets into lowercase.
\item ASCII transliterations of Unicode text (Tweet).
\item Separation of sentence into tokens (words)
\item Stemming of social media tweets using the standard Porter stemmer.
\end{itemize}

\subsection{Statistical Preprocessing}
\label{Stat_prep}
During Disaster-related crises, the social media tweets which belonged to the same context generally submerged with interrelated keywords which summarized the entire tweet (are discussed in Table \ref{tab:keywrds}) and referred in the rest of the paper. These keywords consist of tweets which describe the resource availability status and moreover the need of the resources.
\begin{table*}[!ht]
  \caption{Description of Top five keywords available in Corpus}
  \label{tab:keywrds}
  \begin{tabular}{cll}
    \toprule
    Keyword&Description&Annotation\\
    \midrule
  Question & Request for any resources & $\beta_{need}$\\
  Link & Offer availability of resource using portal & $\gamma_{need}$, $\beta_{availability}$\\
  Hash & Broadcast availability of resource using all popular Hashtags & $\gamma_{availability}$\\
  Mention & Offer some help while publicizing particular community (Political Party etc.) & $\alpha_{availability}$\\
  Number & With Mobile number offer help (Ambulance etc.) & $\alpha_{need}$\\
\bottomrule
\end{tabular}
\end{table*}

\begin{figure}[!ht]
\begin{tikzpicture}[scale=1, transform shape]
\begin{axis}[
    ylabel={Cumulative Frequency of keywords},
    xtick={Question, Link, Hash, Mention, Number},
    legend pos=north west,
    symbolic x coords={Question, Link, Hash, Mention, Number},
]
 
\addplot[
    color=red,
    mark=Mercedes star,
    thick,
    ]
    coordinates {(Question,3)(Link,304)(Hash,950)(Mention,1131)(Number,1759)};
    \addlegendentry{Availability} 

\addplot[
    color=black,
    mark=triangle,
    thick,
    ]
    coordinates {(Question,4)(Link,77)(Hash,249)(Mention,338)(Number,404)};
    \addlegendentry{Need}
  
\end{axis}
\end{tikzpicture}
\caption{Statistical Relationship of top 5 Features among different social media macro community category in entire training corpus}
\label{fig:com_avail_need}
\end{figure}
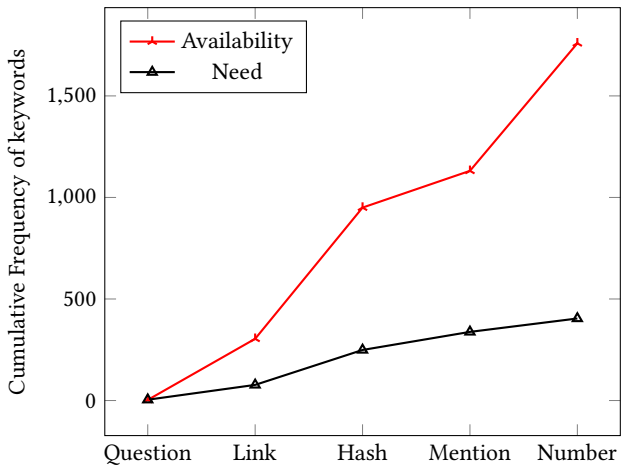
The contribution of these keywords are described in Figure \ref{fig:com_avail_need}, where \textit{Availability Tweets} have the maximum contribution. To have reflection of these tokens in tweet, we generalize some tokens by replacing $\text{?}$ by the token QUESTION, hyperlinks by the token LINK, Hashtag by the token HASH, MENTION token is the replacement of (@username) and lastly, any number of the tweets are replaced by NUMBER.

\section{Methodology}
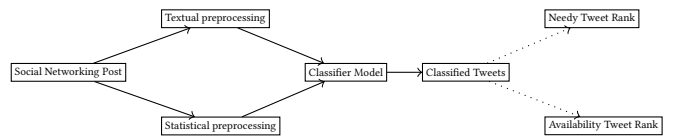
\begin{figure}[!ht]
\begin{tikzpicture}[every node/.style={minimum height={0.5cm},thin,transform shape}, scale=0.47]
\node[draw] (TD) {Social Networking Post};
\node[draw, above right = of TD] (TP) {Textual preprocessing};
\node[draw, below right = of TD] (SP) {Statistical preprocessing};
\node[draw, below right = of TP] (CM) {Classifier Model};
\node[draw, right = of CM] (PC) {Classified Tweets};
\node[draw, above right = of PC] (NRK) {Needy Tweet Rank};
\node[draw, below right = of PC] (ARK) {Availability Tweet Rank};

\draw[->] (TD) -- (TP);
\draw[->] (TD) -- (SP);
\draw[->] (TP) -- (CM);
\draw[->] (SP) -- (CM);
\draw[->] (CM) -- (PC);
\draw[->, dotted] (PC) -- (NRK);
\draw[->, dotted] (PC) -- (ARK);

\end{tikzpicture}
\caption{Architecture and Methodology of Categorical IR}
\label{fig:method_arch}
\end{figure}

Most of the users in Online Social Network (specifically, Twitter) share personal information, random thoughts, opinions/complaints and facts ~\cite{15}. Moreover, with in near-real-time these social network explodes with an activity whenever any calamity takes place \cite{6}. Considering above statements as a hypothesis for this paper, we combine the discussed feature resolution methods to create a fully-automated system, named as Categorical Information Retrieval (IR) and outlined in architecture Diagram \ref{fig:method_arch}. Given the social networking post (tweet, that are associated with the related hashtag), the Categorical IR initiates with the formation of two feature vectors termed as \textit{Textual preprocessing and Statistical Preprocessing}.\\

The formation of these feature vectors are essential as it normalizes the semantics of the given post. These preprocessed posts are transformed into embedding \cite{2} vectors  of size 100 and 5 in length respectively. Once the initialization of entire embedding are done, our classification model (further elaborated in Section \ref{sec:C_Model}) precisely able to categorize any tweets with respect to its relevance with the help of these embeddings. After enforcing these requirements, our Ranking algorithm will precisely rank the classified tweet irrespective of any language within a relevance class.

\section{Classification Model}
\label{sec:C_Model}
Classification model initiates with the formation of words embeddings for every social media tweet using Textual preprocessing with the help of Doc2vec model\cite{2}, well suited for variable length and short length text which outperforms the Word2vec model and probabilistic latent semantic analysis model approach discussed in \cite{7, 20, 26}. Now, each of the word embedding (of size 100 in length) combines another feature vector which was formed using the frequency of each token (discussed in Section \ref{Stat_prep}) available in a tweet. After preprocessing, we have one entire embedding vector of length 105 which will be the final representative of the respective class. \\

Social media tweets in entire training corpus are transformed into 21,00,000 vectors (each of length 105) in which the absence of most similar tweets are ensured by the track organizers. These embeddings will be the input for multilayer preceptron (a feedforward Artificial Neural Network model). Multilayer preceptron model runs upto 300 iterations with an ``ReLu" as a activation function. During the training of model, entire training embedding vectors (2,100,000) were cross-validated with K-folds (k=10), and gives maximum precision and recall in comparison with Naive Bayes, Logistic Regression, SVM even with Random Forest including F-Score.

\section{Category Based Ranking}
Classification model effectively classifies each social media tweet to the respective class and each classified tweet along with the prediction score (termed as $score_{need}$ and $score_{availability}$) are collected separately as depicted in Figure \ref{fig:method_arch}. With the weighted(w)-Ranking Key algorithm, we transform the probability of tweet belong to relevance class with the rank score (termed as $Rank_{score}$). Ranking of tweets based on $Rank_{score}$ helps in prioritizing tweets with most relevant to class considering the presence non relevance tweets.

\subsection{w-Ranking Key Algorithm}
\label{sec:rka}
w-Ranking Key Algorithm proceeds with the initialization of two buckets of top 20 keywords from each relevance class of tweets available in ``Classified Tweet" using Twitter LDA (TLDA) topic modeling model which already outperforms the state-of-the-art model on news corpora discussed in \cite{23,24} and in rest of the paper, we refer these bucket as a $K_{20}-Need$\text{ and }$K_{20}-Availability$ respectively. To provide more weights to tweets which contain keywords as described in Table \ref{tab:keywrds}, we have created two more small buckets labeled them as  $H_{5}-Need$\text{ and }$H_{5}-Availability$ respectively. Moreover, each keyword is classified into most relevance class and annotation of each keyword with respect of class also shown in Table \ref{tab:keywrds}. You may note that each keyword can be assigned to more than one category, as the corresponding keyword could belong to more than one category.\\

For each classified tweet, algorithm makes count the presence of each annotated keyword with $\alpha_1, \alpha_2, \text{ and } \alpha_3$. We pool the relevance class score with the updated relevance class score. Now, if the similarity score between the word available in each tweet and the tokens available in buckets ($H_{5}$ and $K_{20}$ of category) exceeds a certain high confidence threshold C, we exponentially upgrades the pool score with itself (and hence it provides the more weight to keyword which is most relevant). For our similarity metric between keywords, Levenshtein Distance (named after Vladimir Levenshtein, 1965) is preferable among another more obvious candidate in string similarity metrics considering the presence of short text language used in a tweet. Moreover, the algorithm assigns negative weights to the tweet which contains most common keywords available in a most non-relevance tweet such as "quake" and any floating-point number. Entire pseudo code of algorithm is discussed in Appendix A.

\section{Evaluation and Results}
Primarily, the motivation for this paper is to get a quick overview of the current fully-automated state-of-the-art models for disaster reliefs for various countries. We reviewed several Information Retrieval (IR) models but none of them have the better precision for short text and multilingual languages. In this section, we highlight the precision and recall of our proposed fully-automatic IR model for disaster reliefs (obtained with the prediction of class and ranking for 50,000 tweets). 
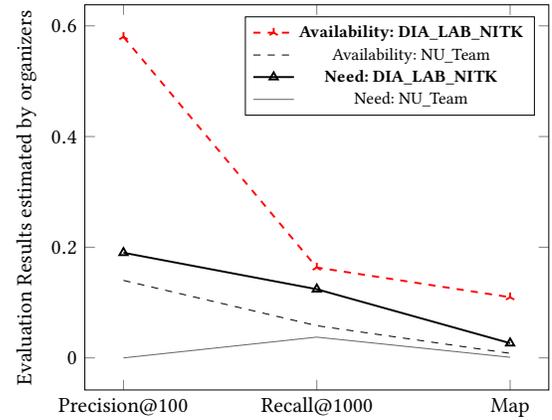
\begin{figure}[!ht]
\begin{tikzpicture}[scale=0.9, transform shape]
\begin{axis}[
    ylabel={Evaluation Results estimated by organizers},
    xtick={Precision@100, Recall@1000, Map},
    legend pos=north east,
    legend style={nodes={scale=0.8, transform shape}},
    symbolic x coords={Precision@100, Recall@1000, Map},
]
 
\addplot[
    color=red,
    mark=Mercedes star,
    dashed,
    thick,
    ]
    coordinates {(Precision@100,0.5800)(Recall@1000,0.1633)(Map,0.1096)};
    \addlegendentry{\textbf{Availability: DIA\_LAB\_NITK}} 

\addplot[
    color=black,
    dashed,
    thin,
    ]
    coordinates {(Precision@100,0.1400)(Recall@1000,0.0582)(Map,0.0082)};
    \addlegendentry{Availability: NU\_Team} 

\addplot[
    color=black,
    mark=triangle,
    thick,
    ]
    coordinates {(Precision@100,0.1900)(Recall@1000,0.1241)(Map,0.0266)};
    \addlegendentry{\textbf{Need: DIA\_LAB\_NITK}}

\addplot[
    color=black!60,
    thin,
    ]
    coordinates {(Precision@100,0.0000)(Recall@1000,0.0375)(Map,0.0011)};
    \addlegendentry{Need: NU\_Team}
  
\end{axis}
\end{tikzpicture}
\caption{Performance evaluation of classification model (including Ranking Key Algorithm) in terms of Precision with top 100 ranked tweets and Recall with top 1000 ranked tweets including comparison of results with another team.}
\label{fig:eval_results}
\end{figure}
The performance evaluation of model shown in Figure \ref{fig:eval_results} while comparing the evaluation results with another Team (NU\_Team) participated in track. Different evaluation metrics (such as precision, recall) and measures were used on top 100 and 1000 tweets (ranked by our algorithm) respectively. Finally with the help of mean average precision results were evaluated while comparing with a gold-standard dataset of Nepal Earthquake.

\section{Conclusions and Future work}
In this paper, we addressed the problem of categorical information retrieval and ranking model for disaster relief while considering the presence of short-text and multilingual languages. In such a scenario, we can expect to have a lot of conversational and sympathy content submerged with the critical information. Moreover, most of the critical information on Online Social Media regarding disaster relief are either in form of short-text or in local regional language. Hence, Our fully-automated \textit{Categorical information retrieval model for disaster relief} integrates textual and statistical preprocessing methods. Through the combination of textual and statistical embeddings we train our model with a multi layer preceptron model with the presence of 76\% Availability Tweets and 23\% of Need Tweets achieved 6.81\% of mean average precision on 50,000 unlabeled disaster relief tweets. Categorical model can be used in evaluation of number available resources with need of resource ultimately it helps in reducing the load on public emergency services and coordinate social effort to save lives. Our experiments on Twitter data from Nepal Earthquake dataset strongly indicates that the proposed fully-automated supervised approach could most often result in improving the accuracy as compared to state-of-the-art approaches and produces the similar result on other dataset even on real time streaming. We plan to extend our approach with deep residual networks and with this strategy it is expected to result in further improvement in precision and recall of architecture. 


\appendix
\section{Appendix}
\begin{algorithm}
\caption{Ranking Key Overview}
\label{alg:generator}
\SetKwProg{generate}{Function \emph{generate}}{}{end}
$\lambda_{Need} \leftarrow \max(score_{need})$\\
$\lambda_{availability} \leftarrow \max(score_{availability})$\\

\generate{(Given Category, Generate $Rank_score$)}{
            \For{each tweet $ \in $ $tweets_{DataBase}$}
            {
                \If{$tweet_{category}$ $ \in $ category}
                {
                    $\alpha_1 \leftarrow$ Number of $\alpha_{category} \in tweet$ \;
                    $\alpha_2 \leftarrow$ Number of $\beta_{category} \in tweet$ \;
                    $\alpha_3 \leftarrow$ Number of $\gamma_{category} \in tweet$ \;
                    \For{each word $ \in $ $(K_{20-category}$ and $H_{5-category})$}
                    {
                        $score_{category}$ $\leftarrow$ $score_{category}$ *  $\lambda_{category}$\;
 
                        \If{ similarity (\{FP or "quake"\}, tweet) > C}
                        {
                            $score_{category}$ $\leftarrow$ $score_{category}$ * -1\;
                        }
                    }
                    $Rank_{score}$ $\leftarrow$ $score_{category}$ \;
                }            
            return $\big( \alpha_1 + \alpha_2 + \alpha_3 \big)$ * $Rank_{score}$ \;
            }
}
\end{algorithm}